\documentclass[a4paper,10pt]{article}
\usepackage{graphics}
\usepackage{amssymb}
\begin{document}

\title{Heun Functions and the energy spectrum of a charged particle on a sphere under magnetic field and Coulomb force}
\author{A.Ralko and T.T.Truong, \\ Laboratoire de Physique Th\'eorique et Mod\'elisation, \\ Universit\'e de Cergy-Pontoise, F-95031, Cergy-Pontoise Cedex, France}

\maketitle

\begin{abstract}
We study the competitive action of magnetic field, Coulomb repulsion and space curvature on the motion of a charged particle. The three types of interaction are characterized by three basic lengths: $l_{B}$ the magnetic length, $l_{0}$ the Bohr radius and $R$ the radius of the sphere. The energy spectrum of the particle is found by solving a Schr\"odinger equation of the Heun type, using the technique of continued fractions. It displays a rich set of functioning regimes where ratios $\frac{R}{l_{B}}$ and $\frac{R}{l_{0}}$ take definite values. \\
Pacs: 03.65.Ge, 73.43.-f
\end{abstract}

\section{Introduction}

The motion of charged particles under constant magnetic field on a sphere has been introduced by F.M.Haldane \cite{arthald} and several authors \cite{artfano,artortiz,artsu} to discuss the fractional quantum Hall effect, and compute several relevant quantities for a system of $N$ independent particles. The sphere is basically a macroscopic support for the motion of a large assembly of particles and its radius $R$ is usually of macroscopic order.

However in recent years, it has been possible to realize spheres of nanostructures ranging from fullerenes size to spherical nanosize objects such as $SiO_{2}$ balls in Opals \cite{artkari,artmig}. The motion of a charged particle on such spheres with constant magnetic field parallel to a given diameter of the sphere has been studied quite extensively \cite{artkim,artaoki,artaris}. The energy spectrum helps to understand some interesting aspects such as orbital magnetism.

In view of future developments, it appears useful to consider the motion of charged particles on a sphere under the action of competiting forces. Besides the curvature of the configuration space and the confining effect of the radial magnetic field, it may be interesting to introduce the Coulomb electric force due to an ion positioned somewhere on the sphere. This situation may arise in the process of fabricating the actual sphere when a charged impurity may slip into the spherical surface and become a center of interaction \cite{artbutuley}. The point is thus to see how such a charged impurity affects the energy spectrum whether or not such effects are in practice relevant. Moreover, the choice of a magnetic monopole generating a radial magnetic field is made to obtain analytical solutions. Since a constant magnetic field parallel to a given diameter does not allow us to find an exact wave function, we replace it with a radial field which competing effects are well described. The strength and range of these effects can be described by three characteristic lengths: 
\begin{itemize}
\item $l_{B}= \sqrt{\frac{\hbar}{eB}}$: magnetic confinement length.
\item $l_{0}= \sqrt{\frac{\hbar^2}{M \kappa e^2}}$: Bohr radius of the hydrogen atom.
\item $R$: radius of the sphere.
\end{itemize}

Here $M$ and $e$ are respectively the mass and the charge of the particle, $B$ the constant radial magnetic field on the sphere and $\kappa = \frac{1}{4 \pi \epsilon_{0}}$ in MKSA unit.

Problems with three characteristic lengths do occur in physics. We may cite the case of one dimensional superconductivity in the presence of twinning defects where electron pairs are moving in constant magnetic field and a finite Dirac Comb \cite{arttruong}  potential. The three characteristic lengths involved are:
\begin{itemize}
\item $l_{c}$: coherence length of the electron pairs.
\item $l_{B}= \sqrt{\frac{\hbar}{eB}}$: magnetic confining length.
\item $d$: period of the Dirac Comb potential.
\end{itemize}

In the present case we look for stationary states of a charged particle on a sphere under a constant radial magnetic field and repelled by a point charge placed on the sphere. This is done by solving the time independent Schr\"odinger equation. The solution will turn out to be expressible in terms of a generalization of Gauss Hypergeometric Function called the Heun Function \cite{artronv}. This new function has been studied by several authors \cite{artronv,arterdel,artsvart} and only one type of Heun function turns out to be relevant for our problem. In fact to fulfill the requirements of quantum mechanics it is necessary to demand that the Heun function be defined all over the sphere and also of square integrable class. This yields a condition on one of the coefficient of the Heun equation, which leads to the quantization of the energy levels.

In section $2$ we present the classical aspects of the dynamics showing that in particular the motion in the $\theta$-direction the energy should be larger than a certain limit. In section $3$ we introduce the quantum treatment which leads to the Heun equation.
 We compute in section $4$ the $R \to \infty$ and $S \to
\infty$ limit, which describes the recovery of the motion in
planar geometry. In this case, the limiting problem is that of the
relative motion of two equally charged particles on plane in
uniform perpendicular magnetic field and under Coulomb repulsion.
Results and comments are left to section $5$.

\section{Classical considerations}

Let $(\theta, \phi)$ be the angular coordinates of a particle of mass $M$ and charge $e$ on a sphere of radius $R$. There exists a constant radial magnetic field $\vec{B}$ pointing outward on the surface of the sphere. $\vec{B}$ is described by the vector potential \cite{arthald}
\begin{eqnarray}
\vec{A} = - \Phi_{0} \frac{S}{2 \pi R} \cot{\theta} \vec{\phi}
\label{eqn01}
\end{eqnarray} 
where $S$ (the half magnetic flux) is the strength of the magnetic monopole generating the magnetic field $\vec{B}$ and $\Phi_{0}=\frac{h}{e}$, the elementary flux unit, and $\vec{\phi}$ the unit vector along the $\phi$ direction on the sphere. In fact we have 
\begin{eqnarray}
\vec{\nabla} \times \vec{A} = \vec{B} = (B_{r}=B, B_{\theta} = B_{\phi}=0 ).
\label{eqn02}
\end{eqnarray} 

We assume in addition the existence of a repulsive Coulomb interaction due to a charge $e$, placed at the North pole of the sphere, which is given by the potential 
\begin{eqnarray}
V(\theta) = \kappa \frac{e^2}{2 R \sin{\frac{\theta}{2}}}.
\label{eqn03}
\end{eqnarray} 

The dynamics of the particle is given by the Hamiltonian:
\begin{eqnarray}
H=\frac{p_{\theta}^{2}}{2 M R^2}+\frac{(p_{\phi}+\hbar S \cos{\theta})^{2}}{2 M R^2 \sin^{2}{\theta}}+\kappa \frac{e^2}{2 R \sin{\frac{\theta}{2}}},
\label{eqn04}
\end{eqnarray} 
where $p_{\theta}$ and $p_{\phi}$ are the canonical conjugate momenta to $\theta$ and $\phi$. Motion at constant energy is defined by 
\begin{eqnarray}
H =E.
\label{eqn05}
\end{eqnarray} 

For the sake of simplicity of the discussion we shall work with dimensionless quantities: $\Pi_{\theta}$, $\Pi_{\phi}$, and $\epsilon$ defined by
\begin{eqnarray}
p_{\theta}= \hbar \Pi_{\theta}, \ \ \ p_{\phi}= \hbar \Pi_{\phi}, \ \ \ E= \frac{\hbar^2}{2 M R^2} \epsilon.
\label{eqn06}
\end{eqnarray} 

Introducing the Bohr radius
\begin{eqnarray}
l_{0} = \frac{\hbar^2}{M \kappa e^2},
\label{eqn07}
\end{eqnarray} 
the equations of energy conservation becomes:
\begin{eqnarray}
 \Pi_{\theta}^2 +\frac{( \Pi_{\phi}+S \cos{\theta})^2}{\sin^{2}{\theta}}+ \frac{R}{l_{0}} \frac{1}{\sin{\frac{\theta}{2}}}= \epsilon.
\label{eqn08}
\end{eqnarray} 

We see the competing roles of geometry and Coulomb repulsion expressed by the ratio $\frac{R}{l_{0}}$.

Moreover in spherical coordinates we have: 
\begin{eqnarray}
B_{r} = (\vec{\nabla} \times \vec{A})_{r} = \frac{-1}{R \sin{\theta}} \frac{\partial}{\partial \theta} \left(\sin{\theta} \frac{\Phi_{0}S}{2 \pi R} \cot{\theta} \right), \nonumber \\
B_{r} = \Phi_{0} \frac{S}{2 \pi R^2} = |\vec{B}| = B.
\label{eqn09}
\end{eqnarray} 
we can then introduce a magnetic length $l_{B} = \sqrt{\frac{\hbar}{eB}}$ and express $S$ as:
\begin{eqnarray}
S= \frac{R^2}{l_{B}^{2}}.
\label{eqn10}
\end{eqnarray} 

Hence the combined interactions can be expressed through the ratios of lengths $ \frac{R}{l_{B}} = \sqrt{S}$ and  $\frac{R}{l_{0}}$.

Note that since $0 < \theta < \pi$, $\sin{\frac{\theta}{2}}$ is always positive and the left hand side of eq.(\ref{eqn08}) is the sum of three positive terms.

Now since $p_{\phi}$ is a conserved quantity ($\phi$ is a cyclic variable) one may set $p_{\phi}= m \hbar$, with $m$ fixed equal to $\Pi_{\phi}$. The orbit in the phase space ($\Pi_{\theta}, \theta$) is a curve of equation:
\begin{eqnarray}
\Pi_{\theta}^{2} + \frac{1}{\sin^{2}{\theta}}\left(m+ \frac{R^2}{l_{B}^{2}} \cos{\theta} \right)^2 + \frac{R}{l_{0}} \frac{1}{\sin{\frac{\theta}{2}}} = \epsilon.
\label{eqn11}
\end{eqnarray} 

Equation (\ref{eqn11}) can be viewed as the motion of a fictitious one dimensional particle in $\theta$-space. The potential function for $0<\theta <\pi$ has a minimum value $\epsilon_0$ which depends on $m$, $\frac{R}{l_B}$ and $\frac{R}{l_0}$, thus the motion is only possible if $\epsilon > \epsilon_0$.
\\
The action for a complete cycle is $J_{\theta}$ given by:
\begin{eqnarray}
J_{\theta} =2 \hbar \int_{0}^{2 \pi} d \theta  \sqrt{\epsilon - \frac{1}{\sin^{2}{\theta}}\left(m+ \frac{R^2}{l_{B}^{2}} \cos{\theta} \right)^2 - \frac{R}{l_{0}} \frac{1}{\sin{\frac{\theta}{2}}}} .
\label{eqn12}
\end{eqnarray} 

Bohr Sommerfeld quantization of the $\theta$-motion by setting:
\begin{eqnarray}
J_{\theta} = h (n + \frac{1}{2}) \ \ \ \ \ \ \ \ \ \ n \in \mathbb{N}
\label{eqn13}
\end{eqnarray} 
yields the quantized energy levels through $\epsilon = \epsilon_{n,m}$ such that: 
\begin{eqnarray}
\int_{0}^{2 \pi} \sqrt{\epsilon - \frac{1}{\sin^{2}{\theta}}\left(m+ \frac{R^2}{l_{B}^{2}} \cos{\theta} \right)^2 - \frac{R}{l_{0}} \frac{1}{\sin{\frac{\theta}{2}}}} d \theta = \pi (n+\frac{1}{2}).
\label{eqn14}
\end{eqnarray} 

Unfortunately the integral can not be evaluated in closed form in terms of simple functions. This aspect will not be treated here.

\section{Quantum motion}

The Hamiltonian of the motion can be suggestively reexpressed in terms of the operators $\vec{\Lambda} = \vec{r} \times (\vec{p}+ e \vec{A})$ as in \cite{arthald}:
\begin{eqnarray}
H= \frac{\vec{\Lambda}^2}{2 M R^2} + \kappa \frac{e^2}{2 R \sin{\frac{\theta}{2}}}.
\label{eqn15}
\end{eqnarray} 
 $\vec{\Lambda}$ has cartesian components:
\begin{eqnarray}
\left( \begin{array}{ccc}
\Lambda_x \\ \Lambda_y  \\  \Lambda_z
\end{array} \right) = \left( \begin{array}{ccc} M_{x} + \hbar S \cos{\phi} \frac{\cos^{2}{\theta}}{\sin{\theta}} \\  M_{y} + \hbar S \sin{\phi} \frac{\cos^{2}{\theta}}{\sin{\theta}} \\  M_{z} -  \hbar S \cos{\theta}
\end{array} \right),
\label{eqn16}
\end{eqnarray} 
with usual definition of $\vec{M}$:
\begin{eqnarray}
\left( \begin{array}{ccc}
M_x \\ M_y  \\  M_z
\end{array} \right) = i \hbar \left( \begin{array}{ccc} \sin{\phi} \frac{\partial}{\partial \theta}+ \cos{\phi} \cot{\theta} \frac{\partial}{\partial \phi} \\ -\cos{\phi} \frac{\partial}{\partial \theta}+ \sin{\phi} \cot{\theta} \frac{\partial}{\partial \phi} \\ -  \frac{\partial}{\partial \phi} \end{array} \right),
\label{eqn17}
\end{eqnarray} 
consequently with 
\begin{eqnarray}
M^2 = - \hbar^2 \left( \frac{\partial^2}{\partial \theta^2} + \cot{\theta} \frac{\partial}{\partial \theta} + \frac{1}{\sin^{2}{\theta}}  \frac{\partial^2}{\partial \phi^2}  \right),
\label{eqn18}
\end{eqnarray} 
one has: 
\begin{eqnarray}
\Lambda^2 = M^2 + \hbar^2 \frac{S^2+ 2 i S \cos{\theta} \frac{\partial}{\partial \phi} }{\sin^{2}{\theta}} - \hbar^2 S^2,
\label{eqn19}
\end{eqnarray} 
and the Schr\"odinger equation for stationary states $ \Psi(\theta,\phi)$ reads:
\begin{eqnarray}
\left(  \frac{\partial^2}{\partial \theta^2} + \cot{\theta} \frac{\partial}{\partial \theta}\right) \Psi(\theta,\phi) +  \frac{ \left(  \frac{\partial^2}{\partial \phi^2}-S^2 - 2 i S \cos{\theta}  \frac{\partial}{\partial \phi} \right)}{\sin^{2}{\theta}} \Psi(\theta,\phi) \nonumber \\
+\left[ S^2 - \frac{R}{l_{0}} \frac{1}{\sin{\frac{\theta}{2}}} + \epsilon \right] \Psi(\theta,\phi)  =0.
\label{eqn20}
\end{eqnarray}

As $\phi$ is a cyclic variable, $p_{\phi}$ is conserved, thus wave function $ \Psi(\theta,\phi)$ can be set of the form  :
\begin{eqnarray}
\Psi(\theta,\phi) = e^{i m \phi} F(\theta),
\label{eqn21}
\end{eqnarray}
with $m \in \mathbb{Z}$ and we get a new equation for $F(\theta)$:
\begin{eqnarray}
F''(\theta) + \cot{\theta} F'(\theta) \nonumber \\
+ \left\{-\frac{m^2+S^2-2mS\cos{\theta}}{\sin^{2}{\theta}} - \frac{R}{l_{0}} \frac{1}{\sin{\frac{\theta}{2}}} + \epsilon + S^2  \right\} F(\theta) = 0. 
\label{eqn22}
\end{eqnarray}

The choice of $m$ as a relative integer guaranties the uniformity of the wave function under $\phi$-rotation. We seek solutions $F(\theta)$ which vanishes at $\theta \to 0$ because of the Coulomb repulsion at the Norh pole and which is square integrable.

We now transform this equation with the change of variable $z= \cos{\theta}$ to bring it back to a known canonical form. $F$ is now a function of $z$, satisfying:
\begin{eqnarray}
(1-z^2) F''(z) -2 z F'(z) \nonumber \\ 
+ \left\{-\frac{m^2+S^2-2mSz}{1-z^2} - \frac{\sqrt{2} R}{l_{0}} \frac{1}{\sqrt{1-z}} + (\epsilon + S^2)  \right\} F(z) = 0 .
\label{eqn23}
\end{eqnarray}

A change of unknown function through the substitution: 
\begin{eqnarray}
F(z) = (1-z)^{\frac{a}{2}}(1+z)^{\frac{b}{2}}P(z),
\label{eqn24}
\end{eqnarray}
where $a$ and $b$ are free parameters to be chosen later, and a last change of variable  $x = \sqrt{\frac{1-z}{2}}$ transforms equation (\ref{eqn23}) into a Heun equation under its canonical form:
\begin{eqnarray}
(x-1)(x-a')P''(x) \nonumber \\
+ \left[(\epsilon'+\delta+\gamma)x-(\epsilon'+a' \delta + \gamma (1+a'))+\frac{a' \gamma}{x} \right]P'(x) \nonumber \\
+ \left[\alpha \beta - \frac{\alpha \beta h}{x} \right]P(x)=0,
\label{eqn25}
\end{eqnarray}
with the relation $\alpha +\beta +1 = \gamma + \delta + \epsilon'$.

This equation is a second-order equation with four regular singularities, located at (x= $0$, $1$, $a'$ and $\infty$). $h$ is called ``auxilliary parameter'' and is significant for our problem.
The solution to this Heun equation is a local Heun function admitting an expansion around $x=0$. Such a solution has a convergent expansion in a domain centered at $x=0$ and for $|x|<1$ and $|x|<|a'|$. The South pole which corresponds to $x=1$ will be a singular point since the expansion is divergent there.

To get eq.(\ref{eqn25}), the coefficients $a$ and $b$ must fulfill the conditions:

\begin{eqnarray}
a^2-b^2 = -4 S m \nonumber \\
a^2+b^2 = S^2 + m^2
\label{eqn26}
\end{eqnarray}

As $F(\theta)$ must not have singularity as $\theta \to 0$ (or $\theta \to \pi$) and equivalently $P(x)$ must be regular for $x \to 0$ (or  $x \to 1$), one must choose $a$ and $b$ positive, i.e. with $S=\frac{R^2}{l_{B}^{2}}$:
\begin{eqnarray}
a = |S-m| \ \ \ \ \ \ \ \ \ \ \ \ \ b=|S+m|.
\label{eqn27}
\end{eqnarray}

The parameters of the Heun equation are then related to the parameters of the problem through:
\begin{eqnarray}
a' &=& -1 \nonumber \\
\gamma &=& 2 a +1 \nonumber \\
\delta &=& \epsilon' = b+1 \nonumber \\
\alpha \beta &=& (a+b)(a+b+2) - 4(\epsilon +S^2) \nonumber \\
\alpha \beta h &=& -4 \frac{R}{l_{0}}
\label{eqn28}
\end{eqnarray}

One possibility to make our solution valid everywhere on the sphere is to reduce the Heun functions down to a Polynomial. But this procedure imposes a condition which makes the radius of the sphere discrete. This is, in general physically not acceptable.

The other possibility is to turn to the procedure of augmented convergence \cite{artronv} which makes use an expansion in terms of Hypergeometric functions \cite{arterdel,artsvart}. Following Ronveaux, we first expand our local Heun function $P(x)$ in terms of hypergeometric functions \cite{artronv}: 
\begin{eqnarray}
P(x) =\sum_{\nu} c_{\nu} y_{\nu}(x) =\sum_{\nu} c_{\nu} F(-\nu, \nu + \delta + \gamma -1; \gamma; x)
\label{eqn29}
\end{eqnarray}
where  $F(-\nu, \nu + \delta + \gamma -1; \gamma; x)$ is a local solution of a hypergeometric equation which matches Heun's solution at the singularities $0$ and $1$ (same exponents at the singularities $0$ and $1$). As $P(x)$ obeys the Heun equation, the $c_{\nu}$ fulfill the three-way recursion relation \cite{artronv}:
\begin{eqnarray}
K_{\nu}  c_{\nu-1} + L_{\nu}   c_{\nu} + M_{\nu}   c_{\nu+1}=0
\label{eqn30}
\end{eqnarray}
where 
\begin{eqnarray}
K_{\nu} = \frac{(\nu + \alpha -1 )(\nu + \beta -1)(\nu + \gamma -1)(\nu + w -1)}{(2 \nu +w-1)(2 \nu +w-2)} \nonumber \\
L_{\nu} = \alpha \beta h +a'  \nu (\nu + w) \nonumber \\
 - \frac{\epsilon' \nu (\nu +w)(\gamma - \delta)+ \left[\nu (\nu+w)+\alpha \beta \right]\left[2 \nu (\nu +w)+ \gamma (w-1) \right]}{(2 \nu + w -1)(2 \nu + w +1)} \nonumber \\
M_{\nu} = \frac{(\nu +1)(\nu +w-\alpha+1)(\nu +w-\beta+1)(\nu + \delta)}{(2 \nu + w +1)(2 \nu + w +2)} 
\label{eqn31}
\end{eqnarray}
with $w= \gamma + \delta -1$.

The behavior of $\left\{ y_{\nu}\right\}$ is given by \cite{arterdel}:
\begin{eqnarray}
lim_{\nu \to \infty} \left|\frac{y_{\nu + 1}}{y_{\nu}} \right| = \left|\frac{1+X}{1-X}  \right|
\label{eqn32}
\end{eqnarray}
where  $X=\sqrt{1-x^{-1}}$.

To determine this limit, we must find the asymptotic behavior of the $\left\{c_{\nu} \right\}$. For $\nu \to \infty$, the coefficients of the recursion relation become:
\begin{eqnarray}
K_{\nu} \to \frac{1}{4} \nu^2, \ \ \ \ L_{\nu} \to -(\frac{1}{2}-a') \nu^2, \ \ \ \ M_{\nu} \to \frac{1}{4} \nu^2,
\label{eqn33}
\end{eqnarray}
and eq.(\ref{eqn30}) tends to the critical equation $\rho^2+2(2a'-1) \rho +1 = 0$ (where $c_{\nu + i}$ has been replaced by $\rho^i$) which admits the roots $\rho_1$ and $\rho_2$ given by:
\begin{eqnarray}
|\rho_1| =\left|\frac{1-A}{1+A}  \right|  \ \ \ \ \ \  |\rho_2| = \frac{1}{|\rho_1|},
\label{eqn34}
\end{eqnarray}
where $A=\sqrt{1-a'^{-1}}$. Erd\'elyi \cite{arterdel} has shown that the convergence of the series occurs in a region determined by
\begin{eqnarray}
\left|\frac{1+X}{1-X}  \right|< \frac{1}{|\rho_1|} = |\rho_2|.
\label{eqn35}
\end{eqnarray} 
This is the so-called augmented convergence phenomena. This inequality implies that the point $x$ lies in the interior of the ellipse $\cal{E}$ with foci at $0$ and $1$. Thus our wave function $P(x)$ is now well defined on each pole of the sphere.
 
  The condition for which the solution of the Heun equation has augmented convergence at $x=0$ and $x=1$ (poles of the sphere) is that the parameter $h$ must be a zero of the continuous fraction \cite{artlorenperron} obtained from the three way recursion relation \cite{arterdel}, i.e.:
\begin{eqnarray}
B_0+\frac{A_1}{B_1+\frac{A_2}{B_2+\frac{A_3}{B_3 +  ...}}} = 0,
\label{eqn36}
\end{eqnarray}
where 
\begin{eqnarray}
A_{\nu} = - \frac{K_{\nu}}{M_{\nu}} \ \ \ \ \textrm{and} \ \ \ \ \ B_{\nu} = \frac{L_{\nu}}{M_{\nu}}.
\label{eqn37}
\end{eqnarray}
Hence for $h=h_n$ we can construct a bona-fide wave function for a stationary state. The set of $h_{n}$ can be obtained by numerical treatment of the continuous fraction. We use the so-called modified Lentz's method \cite{artnum}, calculating by iterations the values of the continuous fraction and stopping when a given  precision is obtained.

The energy spectrum of the problem is of the form: 
\begin{eqnarray}
\epsilon = \epsilon_n = \frac{R}{l_0} \frac{1}{h_n} + \left(\frac{a+b}{2} \right)\left(\frac{a+b}{2}+1 \right)- \frac{R^4}{l_{B}^{4}}
\label{eqn38}
\end{eqnarray}
with $n= 1, 2, ... \infty$.
Now for $S = \frac{R^2}{l_{B}^{2}} $ and $m$ given, all the parameters of the Heun equation are defined and we can find, with classical numerical methods, the solution in the form of the transcendental eq.(\ref{eqn36}). Note that since we are interested by  the energy spectrum, we have replaced $h_n$ in eq.(\ref{eqn36}) by the following expression: 
\begin{eqnarray}
h_n = \frac{4R}{l_0 \left[ 4(\epsilon_n +\frac{R^4}{l_{B}^{4}} )- (a+b)(a+b+2)\right]},
\label{eqn39}
\end{eqnarray}
which allows to compute directly the energy values $\epsilon_n$. This is done in section $5$.

\section{The $R, S \to \infty$ limit}

It is interesting to look at the  $R, S \to \infty$ limit, since it is relevant for many problems in physics. In the fractional quantum Hall effect, a lot of calculations were made in spherical  ( for simplicity) geometry \cite{arthald} then the physical effects studied in planar geometry. For example, the computation of the energy of the neutralizing background, takes a very simple form in the spherical geometry, since it amounts to put a particle of opposite charge at the center of the sphere .  In this limit, the curvature of the sphere vanishes and we expect to recover the plane. We know that $S$ and $R$ are related by equation (\ref{eqn10}), thus when $R \to \infty$, $S$ must go also to infinity. For physical reasons, $S$ must be infinite, because there is a constant perpendicular magnetic field ( i.e. a non-vanishing flux) on the infinite plane. 

Starting with the Schr\"odinger equation on the sphere, after the substitution (\ref{eqn24}), one has:
\begin{eqnarray}
(1-z^2) P''(z) + \left[b-a -(a+b+2)z \right] P'(z)\nonumber \\ 
+\left[\epsilon +S^2 -\left( \frac{a+b}{2}\right) \left( \frac{a+b}{2}+1\right) - \frac{\sqrt{2} R}{l_0 \sqrt{1-z}}\right]P(z) =0.
\label{eqn40}
\end{eqnarray}   

Thus we look for a change of variable to describe the plane. Such a variable can be $\chi = 2 S (1-z)$ and transforms equation (\ref{eqn40}) into:
\begin{eqnarray}
4S \left\{\chi P''(\chi) + \left(a+1-\frac{a+b}{4S} \chi \right)P'(\chi) \right\} \nonumber \\
+ 4S \left\{ \left[ \frac{\epsilon +S^2 -\left( \frac{a+b}{2}\right) \left( \frac{a+b}{2}+1\right)}{4S} -\frac{R}{2 \sqrt{S} \sqrt{\chi}} \right] P(\chi) \right\} \nonumber \\
- \left\{\chi P''(\chi) + 2 \chi P'(\chi) \right\}=0.
\label{eqn41}
\end{eqnarray}   

For $S \to \infty$, we can neglected the last term of this equation and obtain the asymptotic form of the equation:
 \begin{eqnarray}
\chi P_{as}''(\chi) + \left[a+1-\frac{a+b}{4S} \chi \right]P_{as}'(\chi) \nonumber \\ 
+ \left[ \frac{\epsilon +S^2 -\left( \frac{a+b}{2}\right) \left( \frac{a+b}{2}+1\right)}{4S} -\frac{R}{2 \sqrt{S} \sqrt{\chi}} \right] P_{as}(\chi)=0.
\label{eqn42}
\end{eqnarray}   

With the last change of variable $\xi =\sqrt{ \frac{\chi}{2}}$ and the fact that for $S \to \infty$, $a = |S - m| \to S$, $b = |S + m| \to S$  , equation (\ref{eqn42}) reduces to a Biconfluent Heun equation in his canonical form:
\begin{eqnarray}
\xi P_{as}''(\xi) + \left[1 + \alpha - \beta \xi - 2 \xi^2 \right] P_{as}'(\xi) \nonumber \\ + \left[(\gamma - \alpha -2)\xi - \frac{\delta + \beta (1+\alpha)}{2} \right] P_{as}(\xi) =0,
\label{eqn43}
\end{eqnarray}   
where $\alpha$, $\beta$, $\gamma$ and $\delta$ are expressed in terms of $\epsilon$, $S$ and $\frac{R}{l_0}$. 
As shown in \cite{artbazz}, this equation describes the relative motion of two equal planar charged particles in uniform perpendicular magnetic field and under the Coulomb repulsion. So the link between problem and planar problem is established and is locally acceptable when $R, S \to \infty$.

\section{Results and Comments}

\begin{figure}[ht]
\begin{center}
\resizebox{0.8\textwidth}{!}{
\includegraphics{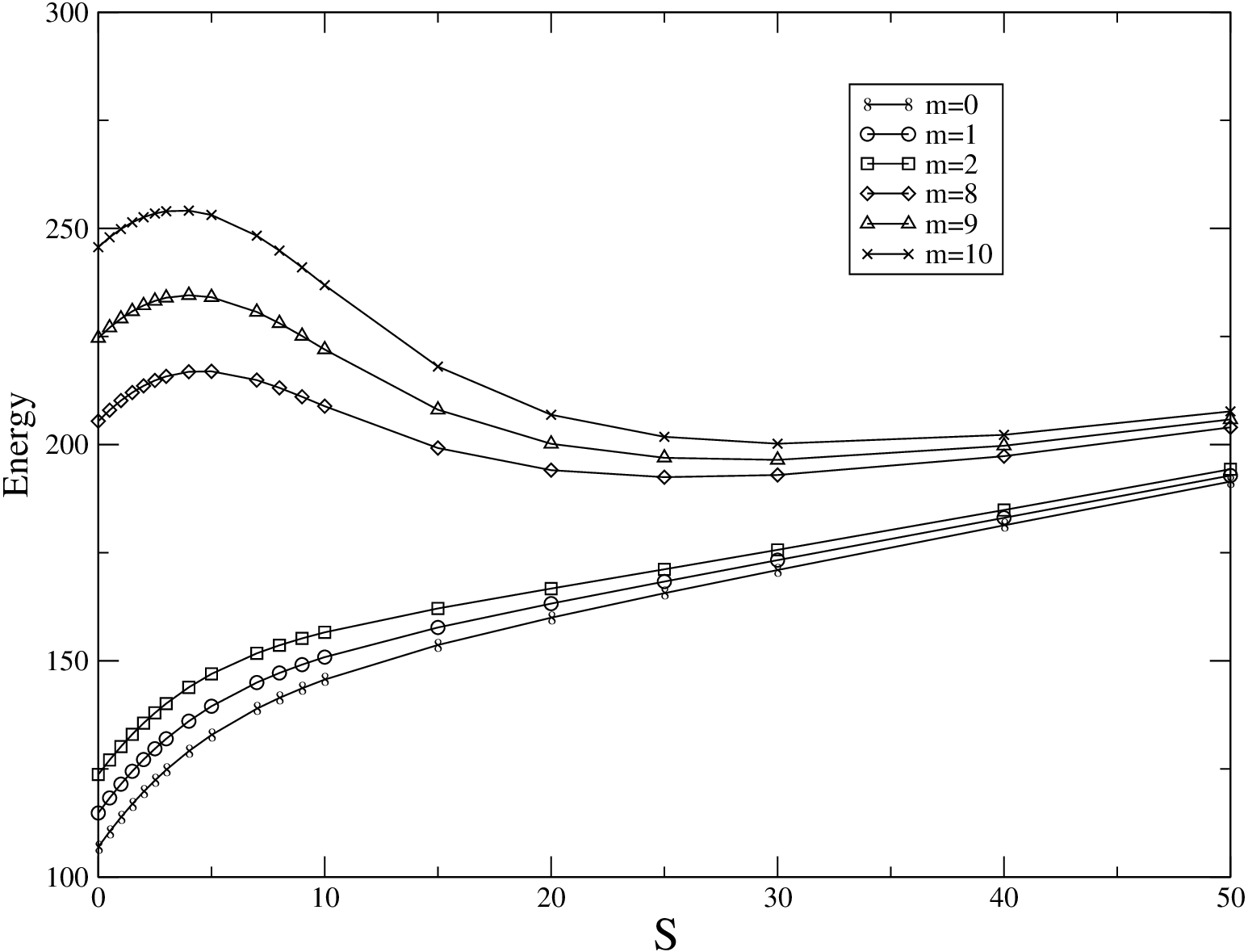}}
\caption{Energy for $n=1$ versus $S$ (proportional to $B$) for positive values of $m$}
\end{center}
\label{fig:01}
\end{figure}

\begin{figure}[ht]
\begin{center}
\resizebox{0.8\textwidth}{!}{
\includegraphics{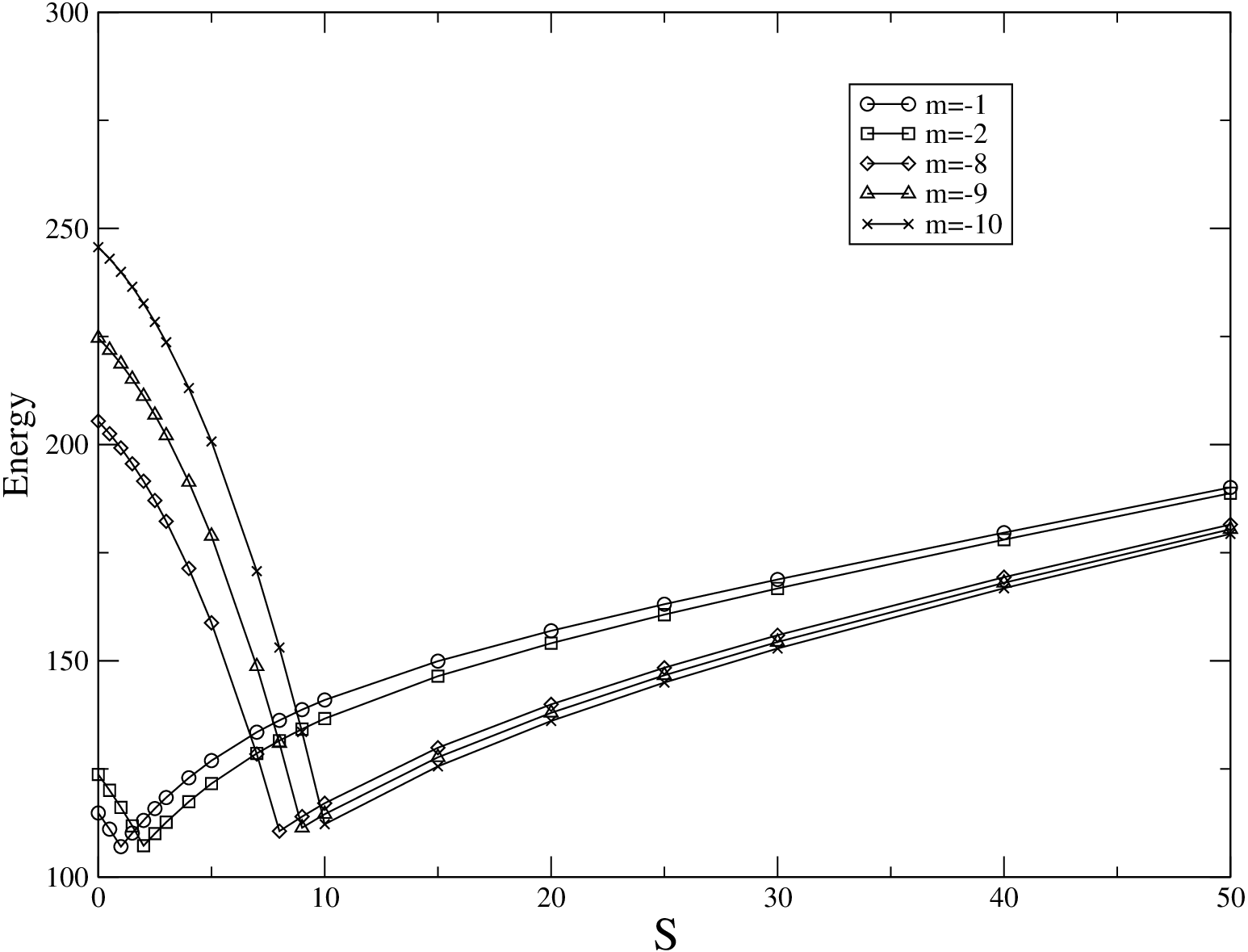}}
\caption{Energy for $n=1$ versus $S$ (proportional to $B$) for negative values of $m$}
\end{center}
\label{fig:02}
\end{figure}

\begin{figure}[ht]
\begin{center}
\resizebox{0.8\textwidth}{!}{
\includegraphics{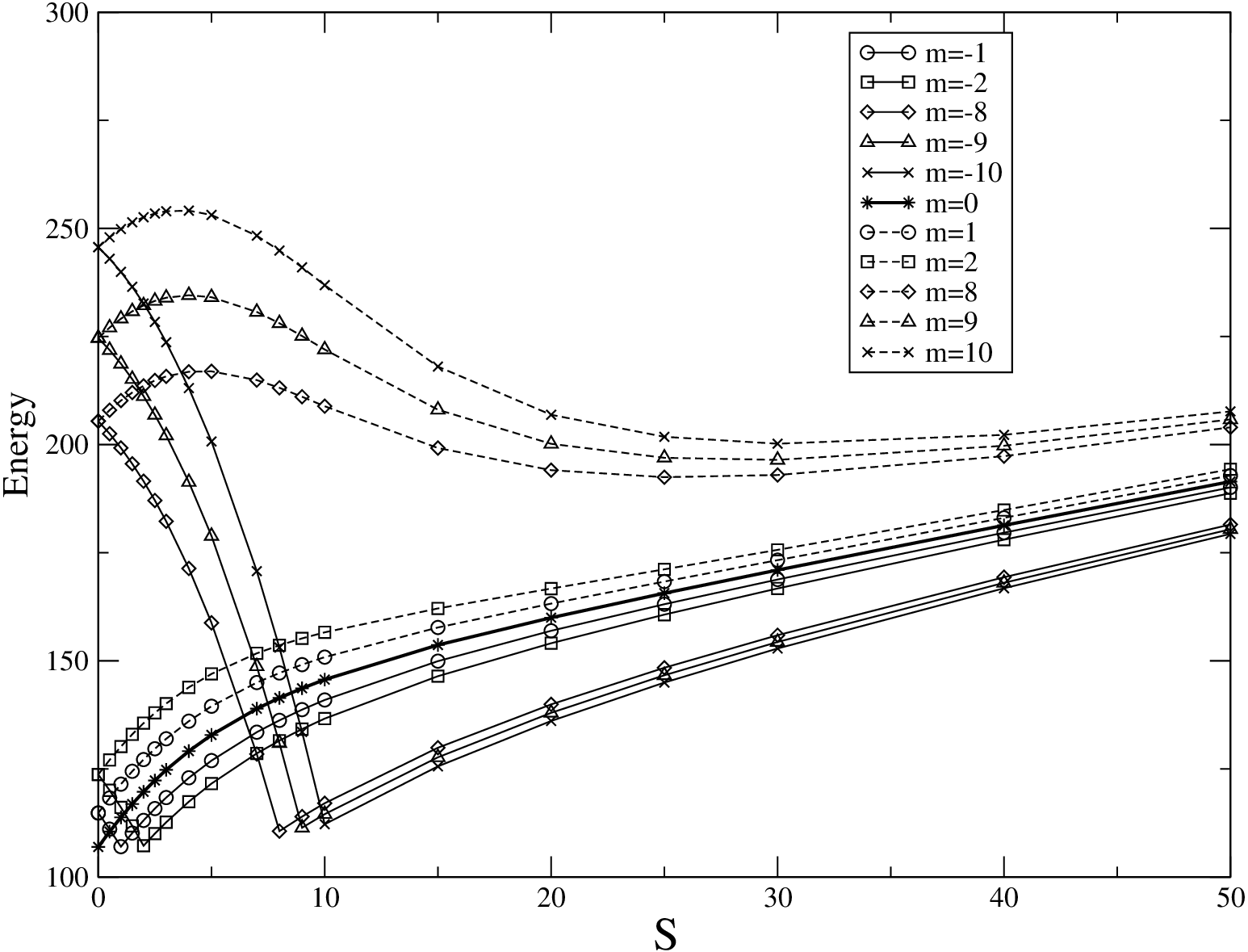}}
\caption{Energy for $n=1$ versus $S$ (proportional to $B$) for $m > 0$ and $m < 0$}
\end{center}
\label{fig:03}
\end{figure}

\begin{figure}[ht]
\begin{center}
\resizebox{0.8\textwidth}{!}{
\includegraphics{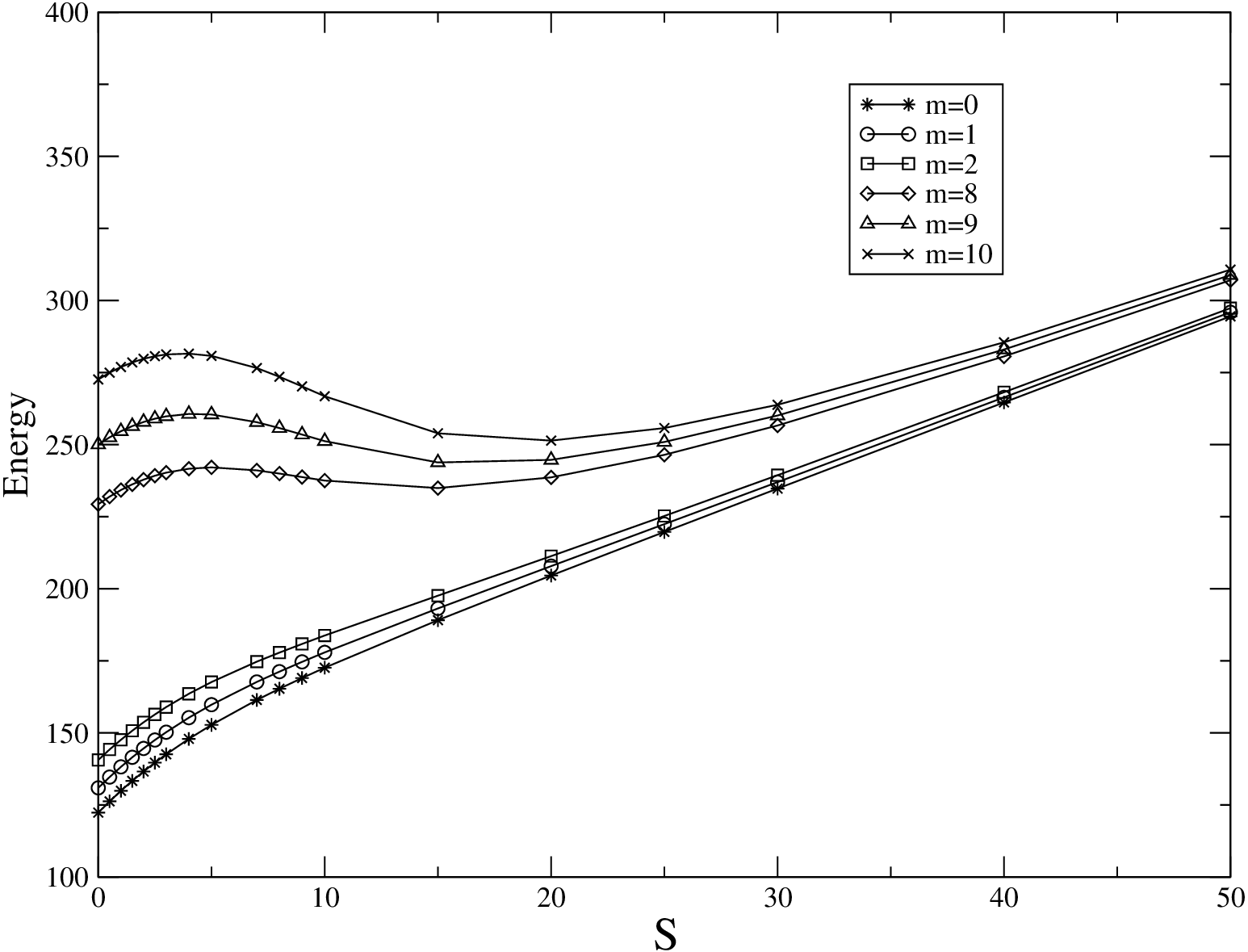}}
\caption{Energy for $n=2$ versus $S$ (proportional to $B$) for positive values of $m$}
\end{center}
\label{fig:04}
\end{figure}

\begin{figure}[ht]
\begin{center}
\resizebox{0.8\textwidth}{!}{
\includegraphics{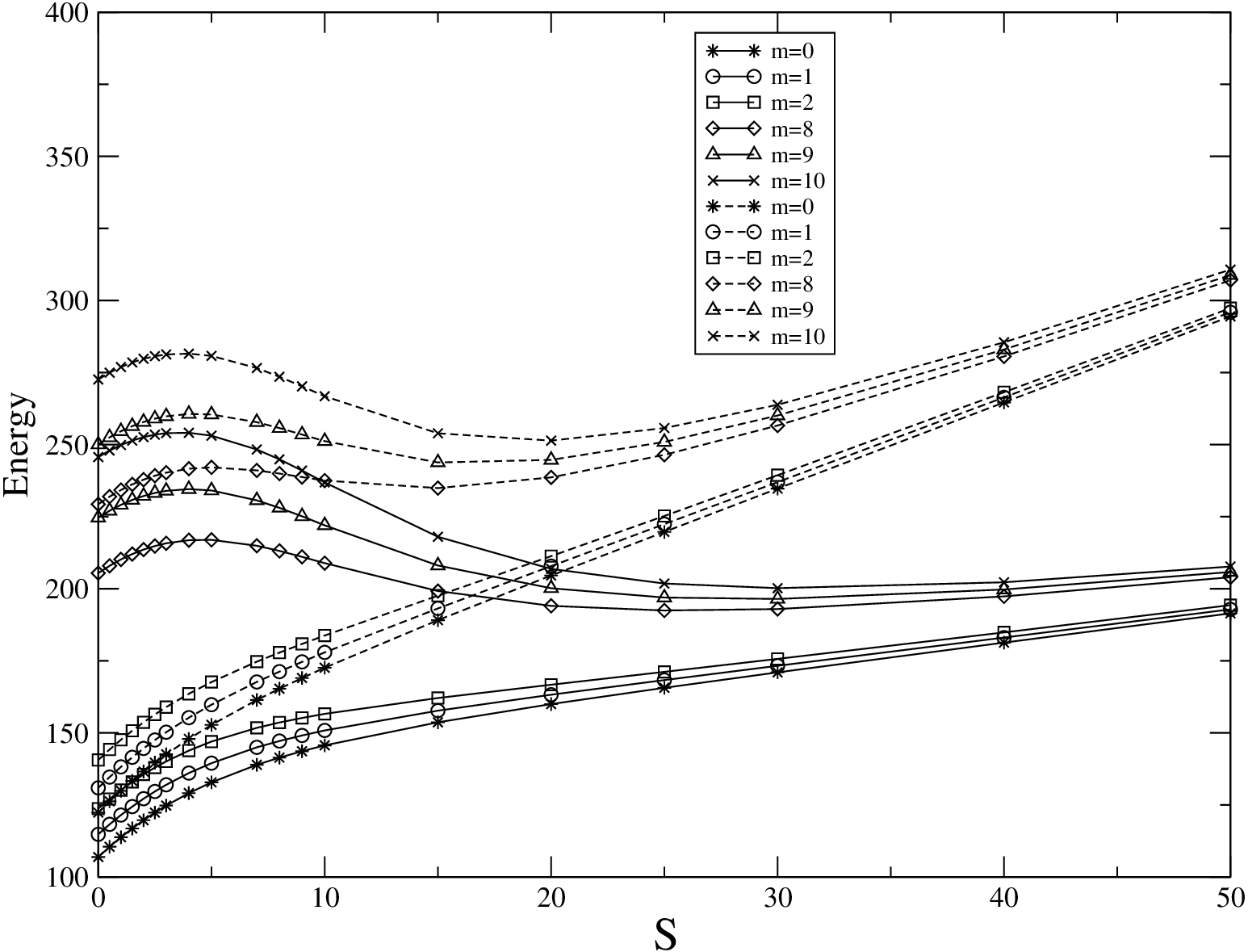}}
\caption{Cross over of the two first levels $\epsilon_{1,m}$ (continuous lines) and $\epsilon_{2,m}$ (broken lines) versus $S$ (proportional to $B$) for positive values of $m$}
\end{center}
\label{fig:05}
\end{figure}

We present the results by giving the two first energy levels $n=1,2$ as a function of $S$ which is basically proportional to the constant magnetic field $B$. The value of the ratio $R/l_{0}$ is fixed for a given radius $R= 100  l_0$ . Figures $1$ and $2$  give the behavior of the ground state energy $\epsilon_{1,m}$ as function of the applied magnetic field for positive and negative values of $m$. 

We see clearly that these values are always bounded below as seen from classical considerations. For negative $m$ values the curves exhibit a sharp discontinuity in the slope of the tangent at some values of the magnetic field ($|m| = S$), there is also a milder discontinuity in the derivative without change of sign for positive values of $m$.
 
In figure $3$, the two situations are put together to display the phenomena of level crossing with respect to the magnetic field.

 The behavior of the level $\epsilon_{2,m}$ for positive values of $m$ is given in figure $4$. 

Finally crossovers between levels $\epsilon_{1}$ and $\epsilon_{2}$ for positive $m$ are displayed in figure $5$.
Note that at $B=0$ we have the spectrum of the charged particle in the presence of pure repulsion centered at the North pole. For strong magnetic field, $B \rightarrow \infty$, we can neglected the Coulomb repulsion, thus the levels tends to the Landau levels as expected.

Higher levels ($n>2$) exhibit similar behavior for $m>0$ and $m<0$ but calculations are more involved so they will not be displayed here.

Let us examine now the limit of vanishing Coulomb repulsion ($l_0 \to \infty$). The energy levels of eq.(\ref{eqn38}) become
\begin{eqnarray}
\epsilon = \left(\frac{a+b}{2} \right) \left(\frac{a+b}{2}+1 \right) - \frac{R^4}{l_{B}^{4}},
\label{eqn44}
\end{eqnarray}
and are no longer given by the $h_n$. They are to be compared with the Landau levels on the sphere given by \cite{arthald}:
\begin{eqnarray}
\epsilon = \left(n+\frac{a+b}{2} \right) \left(n+\frac{a+b}{2}+1 \right) -S^2,
\label{eqn45}
\end{eqnarray}
where $n$ is the Landau level index. We recover then the Lowest Landau level ($n=0$) on the sphere as exected before. It is important to note that Haldane has considered only states with $a=S-m$, $b=S+m$ and $0 \leq m \leq S$.
\\

To our knowledge, this is the first instance where a special class of Heun function, solution of a linear differential equation with four singularities occurs in quantum mechanics. This class of function can handle physical problems which carry three characteristic lengths. The spectrum energy presents different regimes, going from the pure Coulomb problem ($B=0$) to the Landau problem ($B \to \infty$), passing through the well known level mixing regime. It is the main difference with the pure magnetic problem \cite{artkim,artaoki,artaris} where, for weak fields, the energy levels are equally separated (Landau levels) instead of the Coulombic behavior encountered here. Thus the impurity plays an important role in the weak field regime but not in the high field regime. So, the impurity can be ignored for strong magnetic fields, i.e. for $S$ beyond the sharp discontinuity at fixed $m$. The complete study of this wave function and its properties are left for a future work.

\end{document}